\documentclass{article}

\usepackage{PRIMEarxiv}

\usepackage[utf8]{inputenc} 
\usepackage[T1]{fontenc}    
\usepackage{hyperref}       
\usepackage{url}            
\usepackage{booktabs}       
\usepackage{amsfonts}       
\usepackage{nicefrac}       
\usepackage{microtype}      
\usepackage{lipsum}
\usepackage{fancyhdr}       
\usepackage{graphicx}       
\graphicspath{{media/}}     

\title{John Clark's Latin Verse Machine: 19th Century Computational Creativity
 
}

\author{
  Mike Sharples\\
  Emeritus Professor of Educational Technology \\
  The Open University \\
  UK\\
  \texttt{mike.sharples@open.ac.uk} \\
}

\begin{document}
\maketitle

\begin{abstract}
John Clark was inventor of the Eureka machine to generate hexameter Latin verse. He labored for 13 years from 1832 to implement the device that could compose at random over 26 million different lines of well-formed verse. This paper proposes that Clark should be regarded as an early cognitive scientist. Clark described his machine as an illustration of a theory of “kaleidoscopic evolution” whereby the Latin verse is “conceived in the mind of the machine” then mechanically produced and displayed. We describe the background to automated generation of verse, the design and mechanics of Eureka, its reception in London in 1845 and its place in the history of language generation by machine. The article interprets Clark’s theory of kaleidoscopic evolution in terms of modern cognitive science. It suggests that Clark has not been given the recognition he deserves as a pioneer of computational creativity.
\end{abstract}
\keywords{History of computing \and Cognitive simulation \and Natural language generation \and Computational creativity \and hexameter Latin verse \and 19th century cognitive science \and Kaleidoscopic evolution \and Charles Babbage \and Thinking Machine}

\section{Introduction}
In April 1845, handbills were posted in London advertising “Egyptian Hall. The Eureka, a machine for making Latin Verses, exhibited daily, from 12 to 5 \& 7 to 9; with illustrative lectures. Admission 1s."\footnote{The handbill and other documents relating to John Clark are preserved in the archive of the Alfred Gillett Trust, The Grange, Farm Road, Street, Somerset, BA16 0BQ, UK.}

A visitor to the Egyptian Hall in Piccadilly would have seen a wooden cabinet on legs, about the size of a writing bureau. In the course of a lecture about the device, its operator pulled a small rope attached to a lever, and the machine sprang into motion. The interior whirred with clockwork, it played the National Anthem, and through tiny, glazed apertures at the front of the machine letters appeared, producing a line of well-formed Latin, such as:
\begin{quote}
  IMPIA VERBA DOMI CONJUNGUNT CRIMINA MALA
\end{quote}
which roughly translates as “wicked words at home connect evil crimes.” At each pull of the lever, the machine formed a new line. It was capable of generating over 26 million different lines of hexameter Latin verse.

The inventor of this strange machine was John Clark. Clark spent 13 years designing and perfecting the Eureka as a demonstration of what he termed the principle of “kaleidoscopic evolution”. This paper proposes that Clark should be regarded as an early cognitive scientist. A contemporary of Charles Babbage, Clark should be recognized as a pioneer of generative computing.

\section{Background}

The idea of automated production of Latin verse can be traced back at least to the 17th century. In 1677, a mathematician named John Peter published a booklet with the title \textit{Artificial Versifying, or the School-boy’s Recreation: A New Way to Make Latin Verses} [20]. The front cover claimed:
\begin{quote}
Whereby any one of ordinary Capacity, that only knows the A.B.C. and can Count 9 (though he understands not one word of Latin, or what a Verse means) may be plainly taught, (and in as little a time as this is Reading over,) how to make Hundreds of Hexameter Verses, which shall be True Latin, True Verse, and good Sense.
\end{quote}
The booklet depicts six grids, each containing letters. By choosing any six-digit number, the reader could interrogate the tables and generate a line of six Latin words in hexameter rhythm or meter such as “Perfida dicta mihi confirmant somnia multa.” The Appendix shows the six tables in modern lettering, with this author’s attempt to simplify and condense the instructions into one page.

Peter’s tables are an elaborate game, similar to modern Word Search puzzles where words are hidden within a grid of letters. Each table hides nine words that can be found by starting from a letter on the top row (indexed by the chosen digit) and counting forward nine squares to get the next letter and so on. John Peter could have just listed the nine words for each table, but that would have ruined the mystery. 

The tables are ordered according to the strictures of Latin grammar, so the words always produce grammatically correct sentences: Adjective Noun Adverb Verb Noun Adjective. Lastly, Peter carefully chose the words for each table to conform to the hexameter rhythm of classical Latin poetry.  The Artificial Versifying booklet proved popular and led to imitations of the tables over the next century. 

For many centuries, scientists had been devising machines to mechanize arithmetic, calculate tides and determine the positions of planets. Music boxes that played tunes were produced from the late 18th century. An automaton built in the 1770s by the Jaquet-Droz family could hand-write texts, driven by a wheel that coded the text letter by letter [34]. The German inventor Johann Nepomuk Maelzel manufactured music-playing automatons including, in 1805, the Panharmonicon which could imitate orchestral instruments [8]. However, nobody until John Clark had demonstrated a mechanical device to automate the creative process of versification.
\section{John Clark}
John Clark (1785-1853) was born in the village of Greinton near Glastonbury at the southwest of England [2]. He was a first cousin to Cyrus and James Clark, founders of the C\&J Clark shoe company. After leaving school, he joined his uncle’s woollen stocking business, then worked in his family home as a grocer, before setting up a printing firm in Bridgwater, Somerset. He gained a local reputation as a philosopher, author and poet, including writing a continuation of Lord Byron’s Don Juan. Catherine Impey, a local historian, describes John as being “a remarkable character, full of strange idiosyncrasies\ldots He used to wear a sailor blue jacket with brass buttons because it was prettier than a coat, \& was quite regardless of fashion in other ways, wearing his shirt open all down the front.”\footnote{Cited in notes on John Clark by Roger Clark, a descendent of the Clark family, writing in 1950. The notes are held in the archive of John Clark at the Arthur Gillett Trust.}

Clark’s abiding passion was scientific invention. In 1813 he gained a patent for a new method of rubberizing fabric for blow-up air beds which he sold in 1825 to Charles Macintosh who applied this process to waterproofing material and ran a successful raincoat business. Clark was also a skilled clock repairer. For 13 years, from 1832 to 1845, he transferred his knowledge of clockwork to the design, construction and refinement of his Eureka machine for producing hexameter Latin verses [1].
\section{The mechanics of Eureka}
\begin{figure}
\centering
\includegraphics[width=8cm, angle=-90]{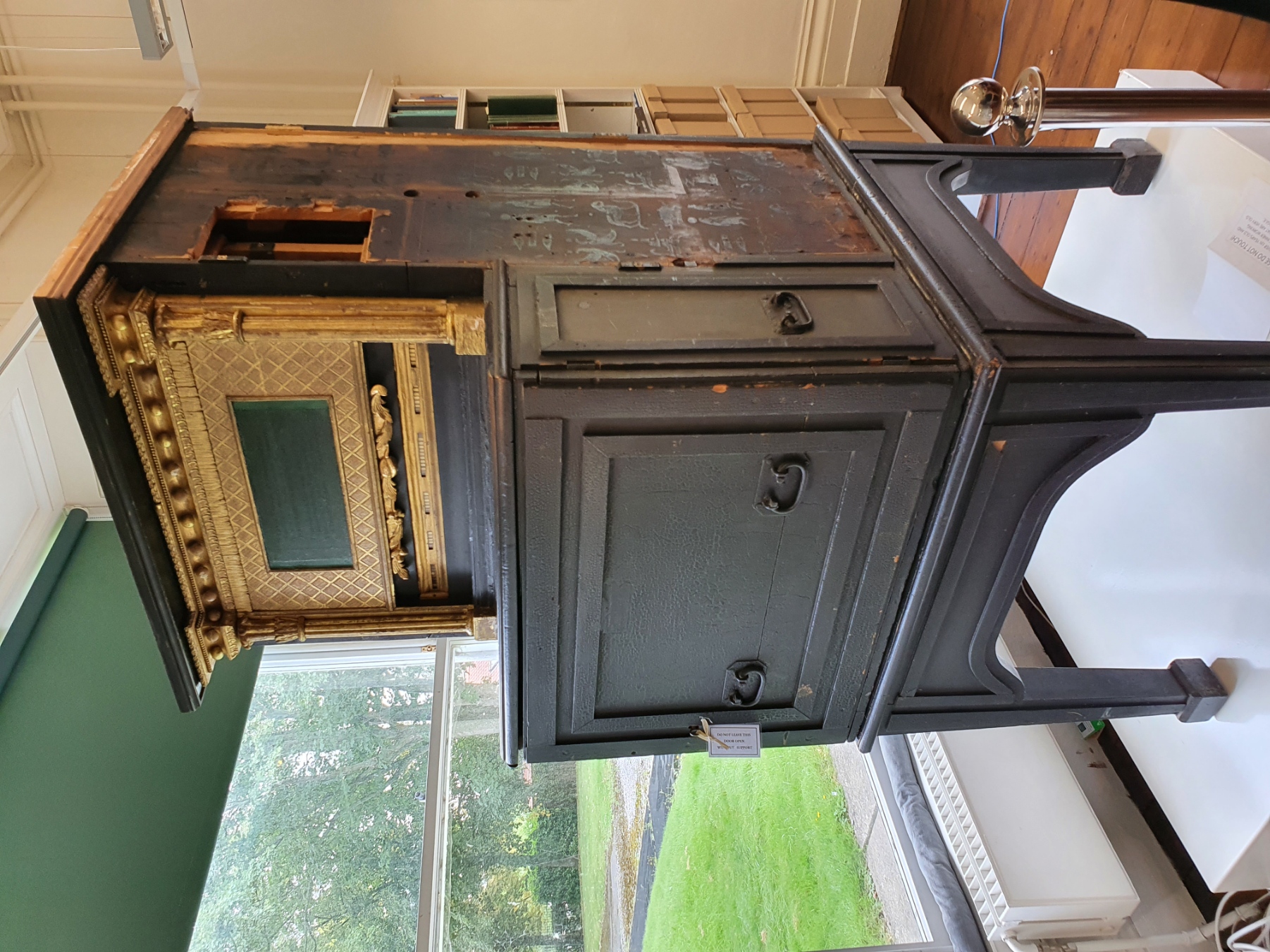}
 \caption{The Eureka Latin verse machine. The verse is displayed on a row of six small windows in the lower strip.}
 \label{fig:fig1}
\end{figure}

In its exhibition version, the Eureka machine was housed in a cabinet on legs, the size of a modern washing machine, painted blue-green with a glossy varnish, with a gilded front panel where a row of small windows displayed the line of verse (Figure \ref{fig:fig1}).

\begin{figure}
\centering
\includegraphics[width=10cm]{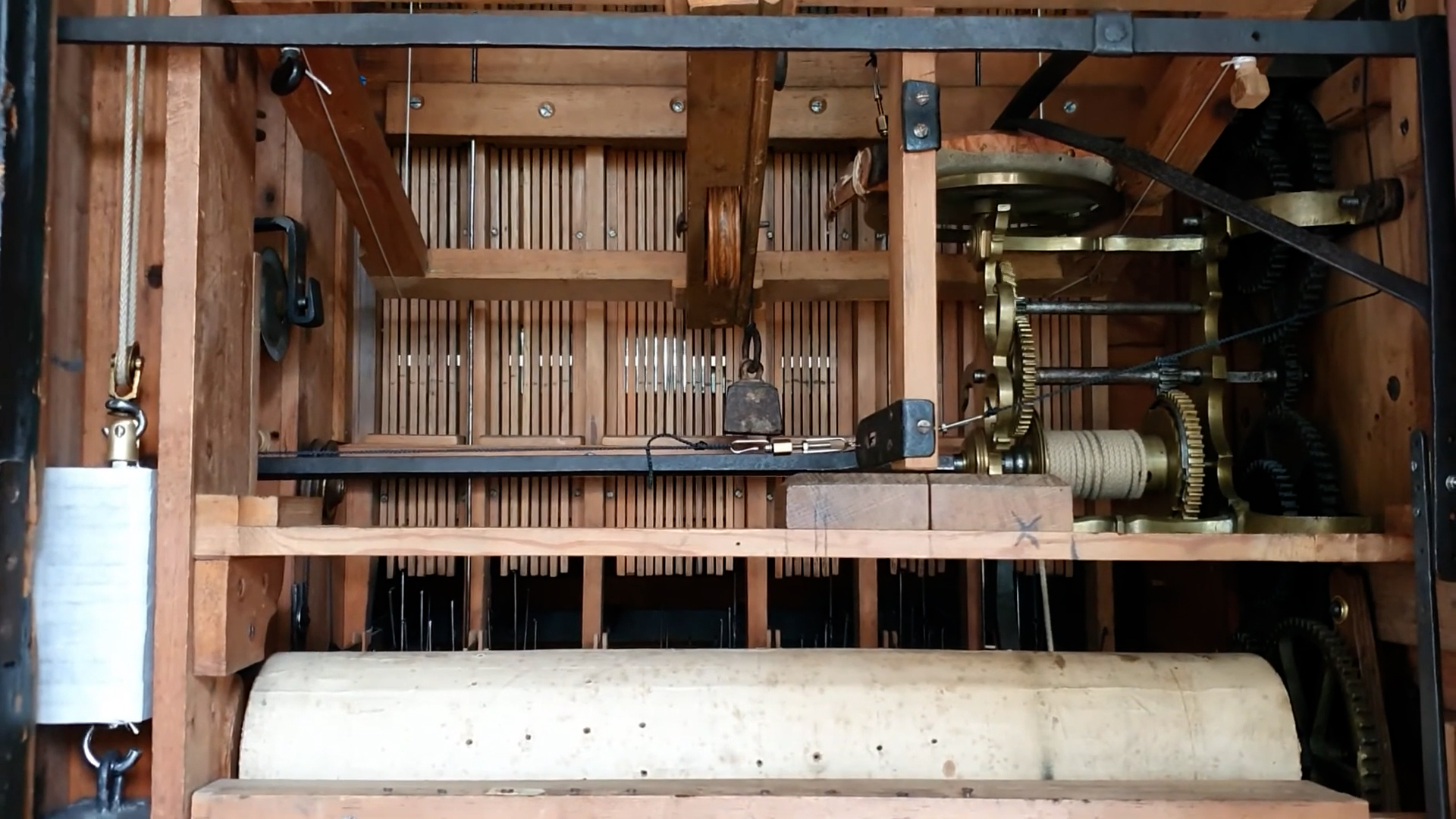}
 \caption{Eureka with the rear door open, showing the weight on the left, driving mechanism on the right, row of wooden staves (shown from behind) and, at the bottom, a large music cylinder that played the National Anthem.}
 \label{fig:fig2}
\end{figure}

Opening the hinged back of the machine reveals\footnote{The present tense is used to describe the construction since, as we discuss later, the Eureka machine still exists in working order.} a clockwork mechanism (Figure \ref{fig:fig2}). A large weight (left of the picture) provides power and a flywheel (top right) governs speed. At the rear of the picture are wooden bars, or staves, that drop onto wires of differing lengths to compose a line of Latin verse. At the bottom is a cylinder, as in a music box, that played the National Anthem to accompany the composing.

\begin{figure}
\centering
\includegraphics[width=8cm, angle=-90]{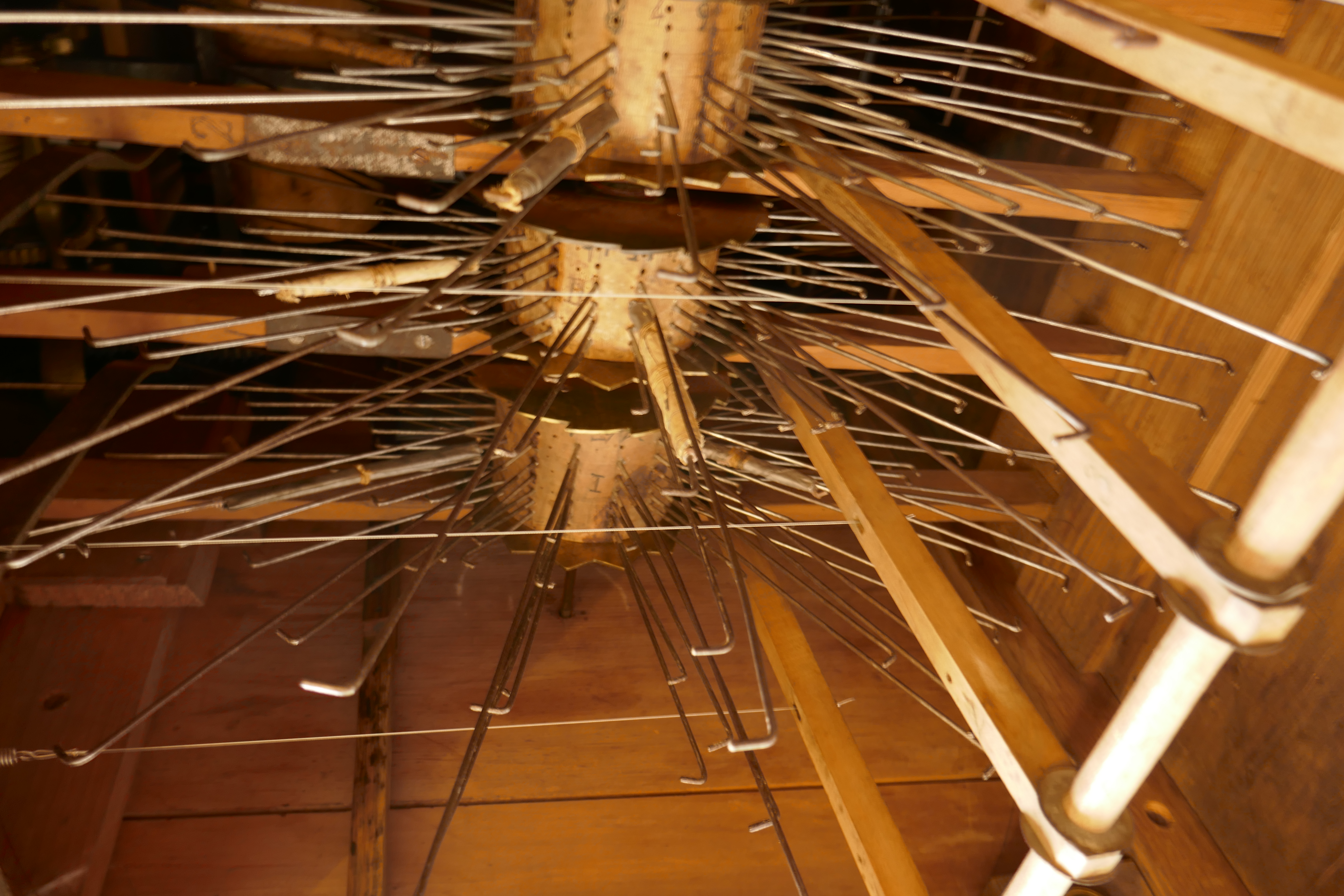}
 \caption{Three of the six drums from the “composing” section of the Eureka machine, showing the rigid wires of differing lengths.}
 \label{fig:fig3}
\end{figure}

\begin{figure}
\centering
\includegraphics[width=6cm]{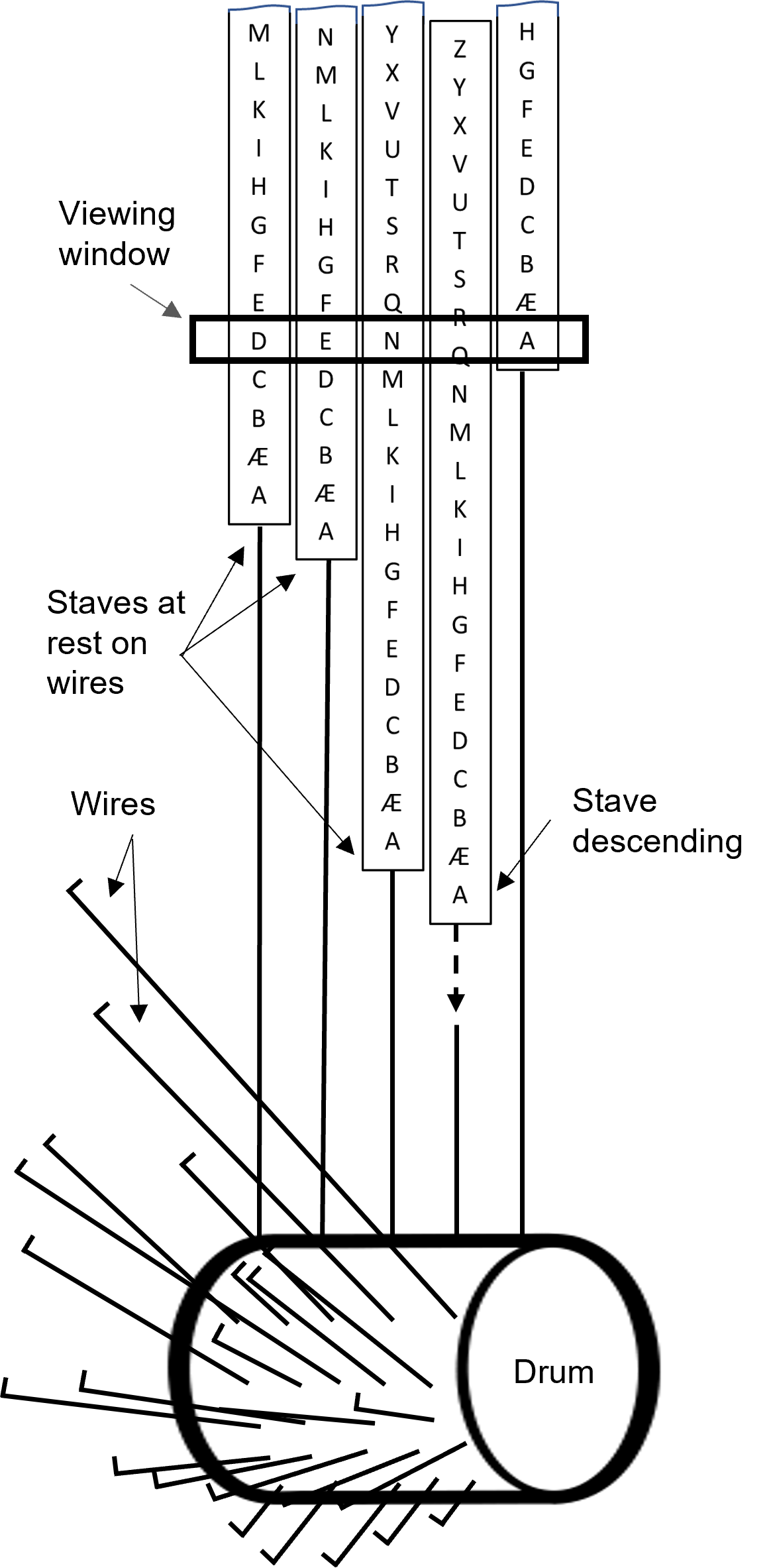}
 \caption{A diagram showing one drum of the Eureka interpreting mechanism, with staves descending onto the wires to display the word DENSA in the viewing window.}
 \label{fig:fig4}
\end{figure}

Below the main drive mechanism is a line of six wooden cylinders, or drums, from which project long rigid metal wires like spines on a porcupine (Figure \ref{fig:fig3}). On each drum, a line of wires, of differing lengths, forms the letters of a single Latin word.
Each wire is bent to a right angle at the end. The drums and wires form the “composing” section of the machine. The drums rotate independently to produce a new line of verse. Behind the clockwork in Figure \ref{fig:fig2} can be seen a line of 47 long vertical wooden staves. On the front of each stave is written a vertical line of letters in alphabetic order:
\begin{quote}
    A Æ B C D E F G H I J K L M N O Œ P Q R S T U V X Y Z
\end{quote}
These form the “interpreting” section. When the machine is set in motion the staves slowly descend to rest on the wires, each displaying a letter of the new verse in the viewing slot. The length of a wire determines a single letter. The longer the wire, the shorter distance a stave has to fall, so the earlier the letter in the alphabet. Conversely, a short wire allows the stave to fall further, producing a letter at the end of the alphabet, such as V or X (Figure \ref{fig:fig4}). By displaying the words gradually, letter by letter as the staves fall, the machine gives the impression it is creating new words, whereas it selects these from a fixed set of words encoded on the drums as lengths of wire. In modern terms, the machine was “programmed” by Clark installing rigid wires of appropriate lengths to form words – the Latin words were literally hard-wired onto the drums.
The machine is wound up by inserting and turning a key on the right of the machine which raises the weight. Moving a lever at the left of the machine sets the clockwork in motion. John Clark describes the mechanism thus [7]:
\begin{quote}
    The entire Machine contains about eighty-six wheels, giving motion to cylinders, cranks, spirals, pullies, levers, springs, ratchets, quadrants, tractors, snails, worm and fly, heart-wheels, eccentric wheels, and star wheels;—all of which are in essential and effective motion, with varying degrees of velocity, each performing its part in proper time and place.
\end{quote}
The machine goes through cycles of interpreting, resetting, and composing as follows\footnote{Adapted from a typewritten “Description of the Mechanism of the Latin Verse Machine” by the conservator Peter Jealous, September 1970. Held in the Clark archive at the Alfred Gillett Trust.}:
\begin{enumerate}
\item The clutch lifts, the control cam moves, and the flywheel begins to revolve.
\item Brake pressure is released, and the machine starts.
\item The flywheel gathers speed, the control frame moves up, and the board that resets the letter staves moves down.
\item The 47 staves independently move slowly down, under gravity. The changing letters can be seen through the viewing windows on the front panel.
\item Each stave comes to rest on a “composing” wire. The front panel shows the appropriate letter.
\item The control cam completes a half cycle and the internal bell rings.
\item The clutch is applied.
\item A complete Latin verse can be read through the apertures on the front panel.
\item The cycle continues, and the letter staves move back up to their start positions.
\item When all the staves are reset and in line, a “drum kick” occurs. Each composing drum rotates a different amount, bringing new rows of wires to the top. Each drum is restrained and the machine is ready for a new cycle of interpreting, resetting and composing.
\item The machine can either be stopped after each cycle or left to continue until the driving weight fully drops. The machine can go through five complete cycles each lasting about a minute before needing to be rewound.
\end{enumerate}
It should be noted that the machine is designed for creativity not calculation. There is no requirement for its movements to be replicable but there is a need to impart arbitrary motion so that each new line of verse is different and unpredictable. Clark included a mechanism to set each drum rotating independently by giving it a rotational kick from a weight-driven pawl that engages a ratchet on the drum to spin it. Then a spring-loaded roller moves to connect with a star wheel on each drum to stop it in a position where the staves can descend onto its line of wires to form a Latin word. The spin mechanism is similar to that seen on mechanical slot machines that were introduced some 40 years later in the late 19th century. It is mechanically (not mathematically) random since the spin of each drum is affected by physical conditions such as temperature and friction.

Clark carefully chose the words coded on each drum. The ordering of the words on the six drums is ADJECTIVE NOUN ADVERB VERB NOUN ADJECTIVE. The machine is designed to always produce a line of dactylic hexameter verse, though it may not always make much sense. He generally chose gloomy words to give a solemn feel to the production, for example:
\begin{quote}
MĀRTĬĂ CĀSTRĂ FǑRĪS PRǢNĀRRĀNT PRŌELĬĂ MŪLTĂ\footnote{The diacritic marks added here indicate the hexameter rhythm: dee dum dum, dee dum, dum dee, dee dee dee, dee dum dum, dee dum.}
\end{quote}
which Clark himself translated as “martial encampments foreshow many oppositions abroad.” 
Table \ref{tab:words} shows the complete set of words for each drum. The physical width of a drum is determined by the longest word coded on it, thus drum 4 is wider than drum 3. The number of possible lines of verse is a permutation of the words on the six drums. Since one word (“foris”) on drum 3 is repeated, the number is 15*16*16*18*19*20 = 26,265,600\footnote{By comparison, John Peter’s system generated 9*9*9*9*9*9 = 531,441 different lines of hexameter verse.}.
\begin{table}
 \caption{Lists of Latin words for each drum on Eureka.}
  \centering
  \includegraphics[width=14cm]{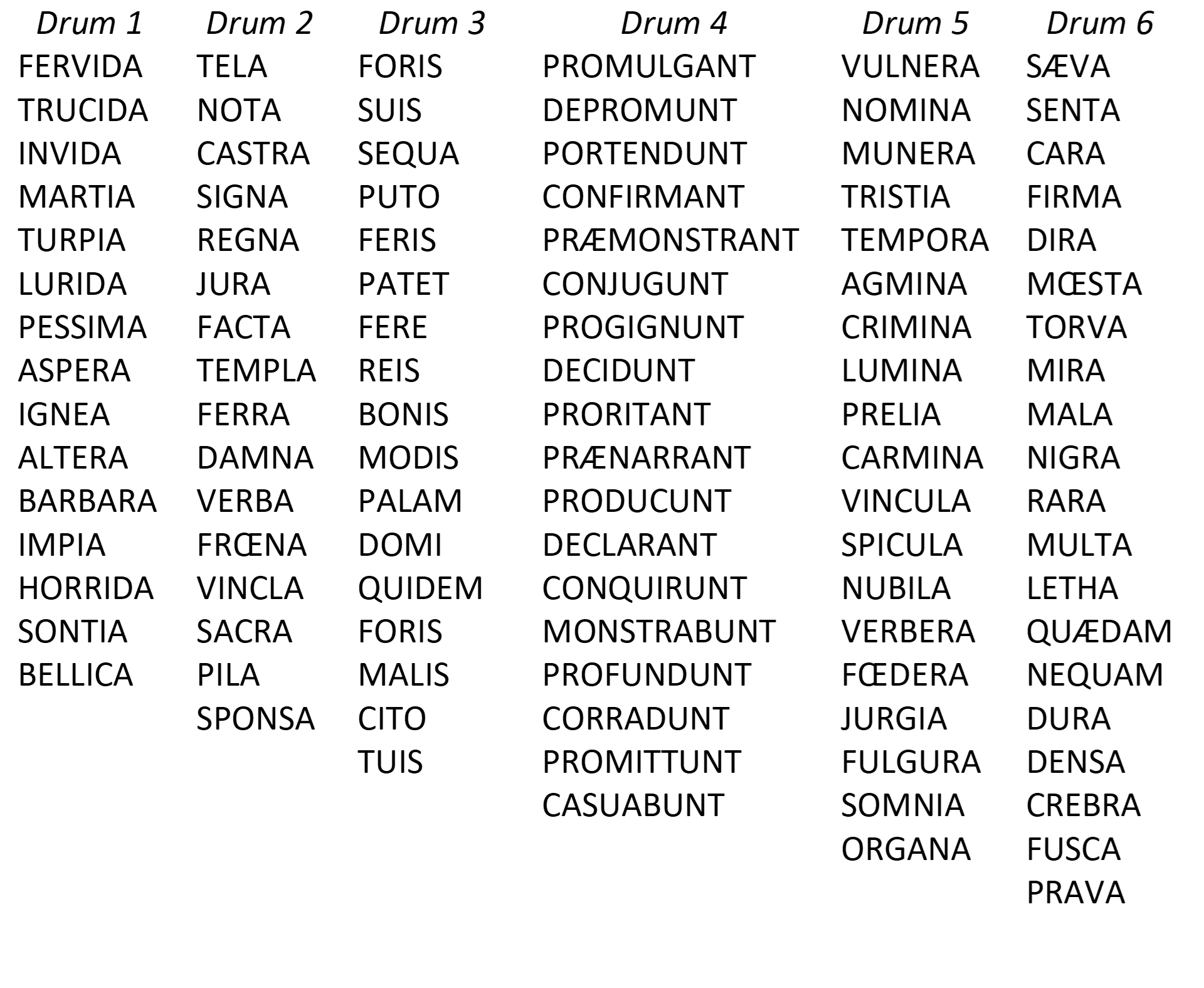}
  \label{tab:words}
\end{table}
Some words, and the general morbid tone, were copied from John Peter but Clark’s vocabulary was larger than Peter’s while retaining his grammar and meter. His choice of Latin words has been criticized for being repetitive and in a few instances incorrect. However, “Faults and shortcomings there may be, but it still stands as a monument to the patience and ingenuity of John Clark” [2]. It was also a practical demonstration Clark’s theory of linguistic creativity.
\section{A 19th Century Cognitive Scientist}
The Stanford Encyclopedia of Philosophy [30] characterizes Cognitive Science as “the interdisciplinary study of mind and intelligence, embracing philosophy, psychology, artificial intelligence, neuroscience, linguistics, and anthropology. Its intellectual origins are in the mid-1950s when researchers in several fields began to develop theories of mind based on complex representations and computational procedures.”

Here, we show how John Clark, working 100 years before the “intellectual origins” of cognitive science, explored a theory of poetic composition that was based on a representation of mental conception realized through a computational mechanism. Although the representation was less complex than those founded on digital computers, it had the hallmarks of a symbolic process theory of mind. Clark named this theory “kaleidoscopic evolution”. 

To put it in historic context, by the early 19th century there was discussion among mathematical scientists as to whether computation machines such as Leibniz’s calculator and Babbage’s Difference Engine could be said to emulate mental processes. In June 1833, Lady Byron (mother of Ada Byron) visited Babbage’s house to view what she called the “thinking machine”. Babbage himself was careful never to ascribe mental powers to his machines [9]. Clark had no such qualms.

In 1837 (five years before the first published description of Babbage’s Analytical Engine) John Clark published a 22-page booklet, which he printed himself, with the title “The General History and Description of a Machine for Composing Hexameter Latin Verses” [5]. A revised version was published in 1848, printed by Frederick Wood of Bridgwater, Somerset [7]. The new version omits a lengthy discussion of poetic forms. It also includes some revealing alterations to wording in the description of the machine and its foundations in a theory of mechanical composition. The account below will refer to the 1848 printing unless stated otherwise. Reference will also be made to a letter Clark wrote to \textit{The Athenaeum} magazine in July 1845, in response to a correspondent from that magazine [6]. In that letter, referring to the exhibition of Eureka in the Egyptian Hall, Clark states:
\begin{quote}
The machine is neither more nor less than a practical illustration of the law of evolution. The process of composition is not by words already formed, but from separate letters. This fact is perfectly obvious, although some spectators may probably have mistaken the effect for the cause—the result for the principle—which is that of kaleidoscopic evolution; and as an illustration of this principle it is that the machine is interesting—a principle affording a far greater scope of extension than has hitherto been attempted. 
\end{quote}
\begin{quote}
The machine contains letters in alphabetical arrangement. Out of these, through the medium of numbers, rendered tangible by being expressed by indentures on wheel work, the instrument selects such as are requisite to form the verse conceived; the components of words suited to form hexameters being alone previously calculated, the harmonious combination of which will be found to be practically interminable.
\end{quote}
The word “indentures” is used in its archaic sense of “depressions” or “furrows” and presumably refers to the lines of wires of differing lengths on the drums. Clark did not use the term “evolution” in its modern definition of evolution of species but the earlier sense of “unfolding” or “coming into being”, applied to a foetus, or in this case a line of verse. Why did he apply the modifier “kaleidoscopic” to this process of composition?

A distinctive feature of cognitive science is the use of technology as a mechanism to understand mental structures and processes. In modern times this is the digital computer; in the mid-19th century, Clark called upon a contemporary scientific instrument, the kaleidoscope.

The kaleidoscope was invented in 1815 by Sir David Brewster as an instrument to create regular patterns from irregular or random arrangements of objects [3]. Clark applied this as a metaphor and mechanism for the creative process of versification. He describes the interior of his machine thus:
\begin{quote}
And in the interior is a large Kaleidoscope, which regularly constructs a geometric figure, the form whereof is ascertained by the falling of numerous probes into its indentures, and thus the Latin verse about to be composed is determined. This action is performed at the commencement of the operation, and is the precise time when the Line of Verse is conceived, previous to its mechanical composition.
\end{quote}
Clark was referring to the kaleidoscopic pattern of wires radiating from the drums that determined how the lettered staves would fall and thus which words would form.

The booklet then refers to “the mind of the machine” to distinguish the “moment of conception” from the mechanical production of the verse (his emphasis below):
\begin{quote}
There is a \textit{certain point of time}, which may be called the \textit{Moment of Conception}, at which instant of time the identical Latin verse, which is \textit{about to be} produced, \textit{is conceived in the mind of the machine}, (if the expression be allowable,) and \textit{that identical verse}, which is then and there conceived, will be mechanically produced and displayed.
\end{quote}
\begin{quote}
This visible display of the Line conceived, is effected by the mechanical movements of the Automaton. But the \textit{conception} of the Line is \textit{not mechanical}, nor can it possibly be rendered a \textit{visible} or \textit{tangible thing}, in being an imagination only, partaking somewhat of the nature of an arithmetic series. In like manner we cannot see or feel human imagination, but we can render it \textit{audible} or \textit{visible} by the mechanical instruments of the tongue or pen.
\end{quote}
\begin{quote}
The Hexameter Automaton bears some affinity to an animated being. It possesses a material and an immaterial part, a corporeal and an incorporeal power. Yet it is a thing far inferior to an animal, inasmuch that it possesses no volition or intention, nor any consciousness of its own existence.
\end{quote}
What follows is an attempt to explain Clark’s conception of kaleidoscopic evolution, with extracts from Clark’s original wording in quotes (his emphasis).

Mechanical reproduction of verse requires two elements: a method of representing the verse (“\textit{to produce the effect by mechanical means}”) and a mechanism to output combinations of well-formed words (“\textit{to arrange the machinery} in a convenient and eligible manner”).

Language can be described as formal sequences of numbers to mechanize composition (“the powers and properties of infinite numbers, applied to poetical composition, etc.”). The language of Latin hexameter verse may be harder for humans to write than everyday English prose (“which act as fetters of confinement to the \textit{writers} of verses, much increasing their difficulties”), but its formal constraints make it easier to implement on a computational device (“have an\textit{ opposite effect} when applied to a \textit{machine}”).

By devising a schema for a line of verse – of the form ADJECTIVE NOUN ADVERB VERB NOUN ADJECTIVE – and selecting words for each category with the appropriate meter, Clark produced a tabular representation (“the foregoing Tabular principle”) where words from each column in sequence could be combined arbitrarily to create a well-formed verse (Table \ref{tab:words}). The process is analogous to a kaleidoscope which, through its careful alignment of mirrors, creates beautiful patterns from arbitrary shapes (“\textit{Forming an indefinite number of regular Geometrical Figures by a Machine}”). Just as a kaleidoscope necessarily creates regular geometric patterns, so the Eureka machine is designed to produce only well-formed verses (“the machine now proposed, \textit{cannot possibly }form other than Hexameter Latin Verses”). 

The verses must not only be grammatically well-formed but also be original (“the Automaton has never repeated any line of verse which it has previously made”). The machine must display creative intelligence, producing verses that make sense (“It is requisite that each Line or Verse shall be \textit{correct}, not only in measure and in accent, but it is also essential that the verse be fraught with \textit{idea} or \textit{intelligence}”).

Clark coded the letters for each word as numbers and implemented these on the Eureka machine as filaments of wire in the six drums, with the length of each wire corresponding to a letter of the Latin alphabet (“Out of these, through the medium of numbers, rendered tangible by being expressed by indentures on wheel work, the instrument selects such as are requisite to form the verse conceived”).

The Eureka machine simulates human creative composition (“the Hexameter Automaton bears some affinity to an animate being”), whereby a Latin scholar who has learned the patterns of hexameter verse can combine these mentally (“an incorporeal power”). The machine mimics this creative process of versification (“at which instant of time the identical Latin verse, which is about to be produced, is \textit{conceived in the mind of the machine}”) and then displays the verse (“\textit{that identical verse}, which is then and there conceived, will be mechanically produced and displayed”). At the “moment of conception” its drums randomly align to produce a novel combination of words (“thus the Latin verse about to be composed, is determined”). The staves then drop down to display the series of letters that form a line of verse. 

The creative mental combination and its mechanical realization are separate processes. The creativity is hidden from view (“we cannot see or feel human imagination, but we can render is \textit{audible} or \textit{visible} by the mechanical instruments of tongue or pen”). Many great ideas have been conceived but not expressed in words (“It is also possible that this supposed \textit{idea} with thousands of others, equally beautiful, and necessarily existing in embryo from eternity to eternity, may possibly \textit{never be disclosed}”). 

The machine can continue to simulate creativity without human assistance (“thousands of Verses may be theoretically conceived, and also mechanically composed and decomposed during the night, or in the intentional absence of all Intelligent Beings, or Spectators: provided that the \textit{weight}, or \textit{power}, which actuates the machine, be continued”).
Clark did not see the Eureka machine as a useless mechanical curiosity but as a working demonstration of his theory of kaleidoscopic evolution, a theory with applications in science and technology. In a “Questions and Answers” section of his booklet Clark asks, “To what uses may the Hexameter Machine be put?” He answers with:
\begin{quote}
It may also be asked of what use is an Acorn? Not much in its present state: but \textit{if planted} and suffered to grow, it may possibly produce an Oak… Thus one invention brings forward another… Every \textit{new thing} is an intellectual \textit{accession}, and every \textit{accession} may, \textit{possibly}, be of important use.
\end{quote}
In his letter to \textit{The Athenaeum} [6], Clark writes “as an illustration of this principle [of kaleidoscopic evolution] it is that this machine is interesting—a principle affording far greater scope of extension than has hitherto been attempted”. In his booklet he envisages automated speech output:
\begin{quote}
A Speaking Automaton\footnote{Presumably a reference to the Euphonia machine, a talking head able to mimic human speech, which was demonstrated in the Egyptian Hall in 1846.}, has lately been completed: it is actuated by a performer, but if it were combined with the Hexameter Machine, it would produce its sentences spontaneously.
\end{quote}
Clark ends the booklet with a brief account of the history of “Androides and Automatic Figures” concluding with:
\begin{quote}
The attention of the present age is deservedly directed to the admirable \textit{Calculating Machine}, designed by Mr. Babbage. This is now in a considerable degree of progress towards completion.
\end{quote}
\section{Reception}
By the time he exhibited Eureka, Clark was aged 60.  His dress and manners must have been out of place among the London intelligentsia. In his memoirs, William Ballantyne Hodgson, a Scottish educational reformer and political scientist describes a visit to the Egyptian Hall to see the Eureka machine [16]:
\begin{quote}
We walked together as far as Piccadilly to the Egyptian Hall, where I saw the Eureka, an instrument for making Latin verse, of which I enclose you a brief account. Had not heard of it before. Barham is exhibiting it just now for the inventor, Clark, whom I also saw. Though I cannot say much for the sense of the verses, there are occasional and recurring errors in quantity\footnote{The “errors in quantity” presumably refers to grammatical errors. Blandford [2] notes that three words (puto, cito and mala) have false quantities.}, and I suspect that the range of the machine is much more limited than is alleged. The inventor spent fifteen years upon it—five more years than are needed to make a boy into a verse-making machine, and still less perfect. Clarke is a strange, simple-looking old man. Babbage said the other day that he was as great a curiosity as his machine.
\end{quote}
The contrast is striking between the suave Cambridge-educated raconteur Babbage and the simply dressed, provincial Clark. By 1845, Babbage had won a Gold Medal from the Astronomical Society and gained over £17,000 of Government grants for his Difference Engine engine to calculate mathematical and astronomical tables, despite never completing a full working machine [28]. Meanwhile Clark was demonstrating his Eureka machine daily in an exhibition hall to a bemused public.

An article in \textit{The Illustrated London News} of July 19, 1845 [32] shows a line drawing of the Eureka cabinet with a factual description of the machine which “has lately been brought to the metropolis, to contribute to the ‘sights of the season’.” Most of the content of that newspaper article is taken from verbatim from Clark’s 1837 booklet.

A letter from “P.A. Nuttall” to The Athenaeum magazine dated June 28, 1845 is less reserved [17]. The Eureka, Nuttall opines, is “little better than a mere puzzle, which any school-boy might perform by a simpler process” since it merely produces six words of a regular Latin grammatical pattern (“the first word is uniformly a dactyl, and an adjective of the neuter plural, the second word a trochee …”). All a school-boy has do it is write Latin words of each type on slips of paper, put them into six piles, then draw out words at random from each pile in sequence to form a Latin verse. “It may be a very curious and instructive amusement,—but nothing more.”

Clark knew that he could have simplified his machine by printing complete words on the drums then rotating each drum to display a line of words through windows, like a 20th-century slot machine. But that would have been less general and creative. As Clark attempted to explain in his response to Nuttall, published in \textit{The Athenaeum} on July 2, 1845 [6], the method he had adopted for Eureka of having the staves fall individually onto the wires is not an effect to entertain the audience but an intrinsic part of its design. He could see how to take advantage of representing the letters physically, not just pictorially. Working with letters not words has “far greater scope of extension than has hitherto been attempted.” Not only could he re-program the machine by adjusting the lengths, but he also envisioned a machine that would automatically interpret the outputs from Eureka to form new verses.

A glimpse of Clarke’s vision for extension to his theory of kaleidoscopic evolution can be seen in a letter he wrote to his sister in August 1845 \footnote{  Letter by John Clark to his sister Sarah (Clark) Metford, August 1845. The letter is held in the Clark archive at the Alfred Gillett Trust.}:
\begin{quote}
I hope to be home soon with a long story to tell. A most astonishing discovery has been made with the Hexameter Machine. It would make so many millions of verses that to produce all will not be done for a century or more, but we have discovered that if we take 100 or so of its productions that these will produce thousands \& millions of other verses, by another machine, \& so on. It has opened a new field of scientific speculation.
\end{quote}
It appears from this fragment that Clark had been speculating about a sequence of machines, each of which takes outputs from the previous machine and combines these in new ways to generate longer and more varied results. 

The exhibition of the Eureka machine in London coincided with what has been called “hexameter mania” [22]. This was a lively debate about the teaching of Classics in schools. In 19th century England, boys in elite schools were required not only to speak and write Latin, but also to compose Latin verse in a manner that supposedly emulated the great Roman poets. By the mid 19th century, this practice had descended into rote learning, with pupils looking up textbooks of verse composition to grind out pastiches of Ovid and Virgil. A machine that could automate this process at the rate of 10,000 verses a week, like a demented schoolboy, added fuel to those commentators who derided the useless rituals of elite schools and were calling for a modern curriculum that embraced the Victorian advances in science and technology. Thus, Clark became embroiled in contemporary debates about the value to the individual and society of writing hexameter Latin verse.

The satirical magazine \textit{Punch} [23] wrote, tongue in cheek:
\begin{quote}
That notable invention, the Eureka, or Latin verse-grinder, was tried yesterday before a committee of young gentlemen from the public schools, who are anxious to have their Latin exercises done with the least possible trouble… Several double-barrelled Eurekas were ordered for Eton, Harrow, and Rugby.
\end{quote}
The debate over the value, if any, of students churning out assignments that could just as well be generated by machine resonates with current concerns about students employing AI language systems to write their essays and assignments [25]. 

The Eureka machine was soon consigned to a footnote in the history of Latin versification. An article for \textit{Chambers’s Edinburgh Journal} in 1850 confuses the Eureka with the Euphonia (a machine to mimic the human voice), stating that “by its aid the most illiterate person could produce thousands of Latin verses” [4].
\section{Legacy}
In the 20th century, the Oulipo movement, originating among writers and mathematicians in France, experimented with automated production of novels and poems but made no reference that we can find to Clark and his verse machine [25]. The best-known mention in literature of a machine to generate verse comes in George Orwell’s \textit{Nineteen Eighty-Four} [19, p. 43]. Orwell depicts the Records Department of the Ministry of Truth where rubbishy newspapers and sentimental songs were “composed entirely by mechanical means on a special kind of kaleidoscope known as a versificator.” Of all the words Orwell might have used to describe a machine for churning out prose and songs he chose “kaleidoscope” and “versificator.” Was that a nod to Clark’s “kaleidoscopic evolution” by a “machine for making Latin verses?”

With the advent of digital computers from the 1950s onwards, many academics and students of computer science (including the author of this paper) experimented with programs to generate poems but without acknowledging the pioneering work of John Clark. In recent years Clark’s work has been re-discovered. In 1963, D.W. Blandford wrote a factual account of Clark and his machine which included a table of words on each drum (Table \ref{tab:words}) [2]. Jason David Hall, Professor of Modern Literature and Culture at the University of Exeter, has written academic papers [10], [11] and a book, \textit{Nineteenth-Century Verse and Technology: Machines of Meter} [12] examining the relation between machine culture and poetic meter in the 19th century. These publications explore the Eureka machine as a producer of metrical verse within the mechanization of science, education, culture, travel and work in Victorian England. The book also examines how 19th-century writers on psychology and physiology saw the human production of meter as an inherently automatic process.

Clark has been recognized by a few researchers in computational creativity as a pioneer of mechanized composition. Douglas Summers Stay, in his book \textit{Machinamenta} has commented on the connection between the Eureka and the kaleidoscope as a machine to produce patterns by random combination. The problem with both, he notes, is how to make a machine that is not limited by the inventor’s initial choice of settings, but can grow in ability over time [27, p.5]. Other mentions of Eureka in the literature on computational creativity include [12], [33] and [25].

As for the Eureka machine itself, it is not clear where it went immediately after Clark’s death in 1853 \footnote{The history of the Eureka machine is drawn from [1].}. Around 1856 it was housed in the Counting House of the Clarks Shoe Company in Street, Somerset. Then, in 1889 or 1890 it was moved to the geology Museum at Crispin Hall in Street, to the concern of Alfred Gillett, donor of the geological collection, who thought it let down his exhibits. The last person capable of operating the machine during that period was John Aubrey Clark, son of the founder of the shoe company. When the Geology Museum dispersed in 1948, the Eureka was moved back to the Clarks factory. It was restored to working order in 1950 by Leslie W.M. Husbands, a local clockmaker, and Frederick Berry, the typewriter and sewing machine engineer at Clarks. Then it was moved to the company Records Office. By 1963 it was in the Clarks company museum. In 1970-71 it was renovated for a second time by the Clarks Special Projects Engineer Peter Jealous, in collaboration with Husbands, and in 1979 it was housed in the company’s Shoe Museum. By 1996, no longer working, it was put into storage. It stayed there until 2015 when it went to Devon for extensive restoration by accredited conservators Richard Jaeschke who conserved the casework and exposed the gilding and verse on the front of the machine, and Neil Bollen who conserved the mechanical parts and restored the machine to a gentle working order. The restoration was part of a collaboration between the University of Exeter and the Alfred Gillett Trust, funded by the Arts and Humanities Research Council, to understand, document and restore Eureka. Eureka now rests on a plinth at the Alfred Gillett Trust, Street, Somerset as a centerpiece of its collection of artefacts connected with the Clarks shoe company \footnote{Eureka can be accessed at the Alfred Gillett Trust in Street, Somerset, by appointment. Researchers can also view archives and collections related to Clarks shoe company, the family, and Street whilst a new museum is being developed.}.
\section{Conclusion}
John Clark does not fit comfortably into the narrative of pre-electronic computing, where urbane and sometimes irascible mathematicians based in the great European cities of Paris and London developed machines of intricate engineering to solve problems in accounting, astronomy and navigation. Clark was an outsider. He never studied at university, he was a self-taught poet and philosopher, he lived most of his life in the market town of Bridgwater on the Southwest corner of England. He began work on the Eureka machine in 1830, when he was forty-five. When, at the age of 60, he arrived in London it was to exhibit his machine at the Egyptian Hall which by that time was noted more for its popular entertainment than display of scientific discovery. 

Clark himself, in the final section of his booklet [7], places his device among “The construction of different species of Androides and Automata [that have] occasionally engaged the attention of Mechanists, of all nations” including “a fine automaton, representing a Bengal Tiger. The deep sounds of its roaring are admirably produced by the organ pipes of its internal structure.” It is hardly surprising given his background, his reception in London (“as great a curiosity as his machine”), and the criticism of Eureka for producing gloomy and occasionally ill-formed \footnote{“From the metrical point of view the inventor would have done better to have produced lines of five words each, arranged perhaps in the order of the so-called Golden Line with nouns and adjectives carefully balanced” [2, p.77].} Latin verses, that John Clark has not been included in histories of computing and artificial intelligence (e.g., [21], [29], [18]). The comprehensive book on \textit{The Origins of Digital Computers} [24] mentions Clark only in the bibliography.

However, the enterprise in which Clark was engaged, to understand and simulate the creative production of language, has been a recurring interest of computer scientists from Christopher Strachey in the 1950s [26], through the extensive work of Sheldon Klein and colleagues at the University of Wisconsin to automate the writing of novels [14] and recent work on computational poetry (see [15] for a review), to the investment by companies including Microsoft, Google, Meta and Baidu in pre-trained AI models for language generation.

This author prompted the GPT-3 Transformer language model from OpenAI to generate original Latin verse. It responded in seconds with well-formed Latin but it failed in rhyme and meter. Clark would have been intrigued and amused.
\section{Acknowledgements}
My sincere thanks go to Karina Virahsawmy, Assistant Curator, Alfred Gillett Trust, for allowing me access to the John Clark archive and giving a personal demonstration of the machine, to Neil Bollen, conservator, for his patient and detailed responses to my queries about how the machine functions, and to Rafael Pérez y Pérez, co-author of \textit{Story Machines: How Computers Have Become Creative Writers}, for setting me on the journey to Eureka.
\section{References}
\begin{description}
\item[][1] C. Berry, “LVM Latin Verse Machine c 1830 – c1845: Collection Description compiled by Charlotte Berry”, The Alfred Gillett Trust, Street, Somerset, United Kingdom, 2014.

\item[][2] D. W. Blandford, “The Eureka,” \textit{Greece \& Rome}, vol. 10, no .1, pp. 71–78, 1963.

\item[][3] D. Brewster, \textit{The Kaleidoscope: Its History, Theory, and Construction with its Application to the Fine and Useful Arts}, 2nd ed. London, UK: John Murray, 1858.

\item[][4] Chambers’s Edinburgh Journal, “Latin versification for the million,” \textit{Chambers’s Edinburgh Journal}, no. 13, p. 205, 1850.

\item[][5] J. Clark, \textit{The General History and Description of a Machine for Composing Hexameter Latin Verses}. Bridgwater, Somerset, UK: John Clark, 1837.

\item[][6] J. Clark, “Letter to \textit{The Athenaeum}”, \textit{The Athenaeum}, no. 923, Jul. 5, 1845, p.p. 669–670.

\item[][7] J. Clark, \textit{The General History and Description of a Machine for Composing Hexameter Latin Verses}. Bridgwater, Somerset, UK: Frederick Wood, 1848.

\item[][8] A. Engberg-Pedersen, “The sense of tact: Hoffmann, Maelzel, and mechanical music,” \textit{The Germanic Review: Literature, Culture, Theory}, vol. 93 no. 4, pp. 351–372, 2018.

\item[][9] C. D. Green, “Was Babbages’s Analytical Engine intended to be a mechanical model of the mind?,” \textit{History of Psychology}, vol. 8, no. 1, pp. 35–45.

\item[][10] J. D. Hall, “Popular prosody: Spectacle and the politics of Victorian versification,” \textit{Nineteenth-Century Literature}, vol. 62, no. 2, pp. 222–249, 2007.

\item[][11] J. D. Hall, “Mechanized metrics: From verse science to laboratory prosody, 1880–1918,”\textit{ Configurations: A Journal of Literature, Science and Technology}, vol. 3, no. 17, pp. 285–308, 2009.

\item[][12] J. D. Hall, \textit{Nineteenth-Century Verse and Technology: Machines of Meter}, Basingstoke, UK: Palgrave Macmillan, 2017.

\item[][13] M. Hrešková and K. Machová, “Haiku poetry generation using interactive evolution vs. poem models,” \textit{Acta Electronica et Informatica}, vol. 17, no. 1, pp. 10–16, 2017.

\item[][14] S. Klein et al., “Automatic novel writing: A status report,” The University of Wisconsin, \textit{Technical Report \#186}, Computer Sciences Department, The University of Wisconsin, 1973.

\item[][15] C. Linardaki, “Poetry at the first steps of Artificial Intelligence,” \textit{Humanist Studies \& the Digital Age}, Vol. 7, No. 1, 2022, doi: 10.5399/uo/hsda/7.1.6. 

\item[][16] J. M. D. Meiklejohn, Ed., \textit{Life and Letters of William Ballantyne Hodgson}, Edinburgh, UK: David Douglas, 1883.

\item[][17] P. A. Nuttall, “Letter to \textit{The Athenaeum},” \textit{The Athenaeum}, no. 922, Jun. 28, 1845, p. 638.

\item[][18] G. O’Regan, \textit{Introduction to the History of Computing: A Computer History Primer}. Springer International Publishing, 2016.

\item[][19] G. Orwell, \textit{Nineteen Eighty-Four} (Penguin Classics edition, 2000), Penguin Random House, 1949.

\item[][20] J. Peter, \textit{Artificial Versifying, or the School-boy’s Recreation}. London, UK: John Sims, 1677.

\item[][21] V. Pratt, \textit{Thinking Machines: The Evolution of Artificial Intelligence}. Oxford: Basil Blackwell, 1987.

\item[][22] Y. Prins, “Metrical translation: nineteenth-century Homers and the hexameter mania,” in \textit{Nation, Language, and the Ethics of Translation}, M. Wood and S. Bermann, Eds., Princeton, USA: Princeton University Press, 2005, p. 229–256.

\item[][23] \textit{Punch}, vol. 9, July to December 1845, London: Punch, p. 20.

\item[][24] B. Randell, Ed., \textit{The Origins of Digital Computers: Selected Papers}, 3rd ed., Berlin: Springer-Verlag, 1982.

\item[][25] M. Sharples and R. Pérez y Pérez, \textit{Story Machines: How Computers Have Become Creative Writers}, Routledge, 2022.

\item[][26] C. Strachey, “The ‘thinking’ machine,” Encounter, Oct. 1954, pp. 25-31, 1954.

\item[][27] D. Summers Stay, \textit{Machinamenta: The Thousand Year Quest to Build a Creative Machine}, Machinamenta Publishing, 2011.

\item[][28] D. Swade, \textit{The Cogwheel Brain}, London, UK: Abacus, 2000.

\item[][29] D. Swade, “Pre-electronic computing,” in \textit{Dependable and Historic Computing}, C.B. Jones and J.L. Lloyd, Eds., Berlin, Germany: Springer, 2011, pp. 58–83.

\item[][30] P. Thagard. “Cognitive Science.” \textit{Stanford Enyclopedia of Philosophy}, https://plato.stanford.edu/entries/cognitive-science/ (accessed Oct. 5, 2022).

\item[][31] The Athenaeum, “Announcement,” \textit{The Athenaeum}, no. 922, Jun. 28, 1845, p. 635.

\item[][32] The Illustrated London News, “The Eureka,” \textit{The Illustrated London News}, Jul. 19, 1845, p. 37.

\item[][33] V. Todorovic, and D. Grba, “Wandering machines: narrativity in generative art,” \textit{J. Sci. and Technol. of the Arts}, vol. 11, no. 2, pp. 50-58, 2019.

\item[][34] A. Voskuhl. “Producing objects, producing texts: accounts of android automata in late eighteenth-century Europe,” \textit{Studies in Hist. and Phil. of Sci}. Part A, vol. 38, issue 2, pp, 422–444, 2007.
\end{description}
\newpage
\appendix
\section{Appendix}
\begin{figure}[h]
\centering
\includegraphics[width=16cm]{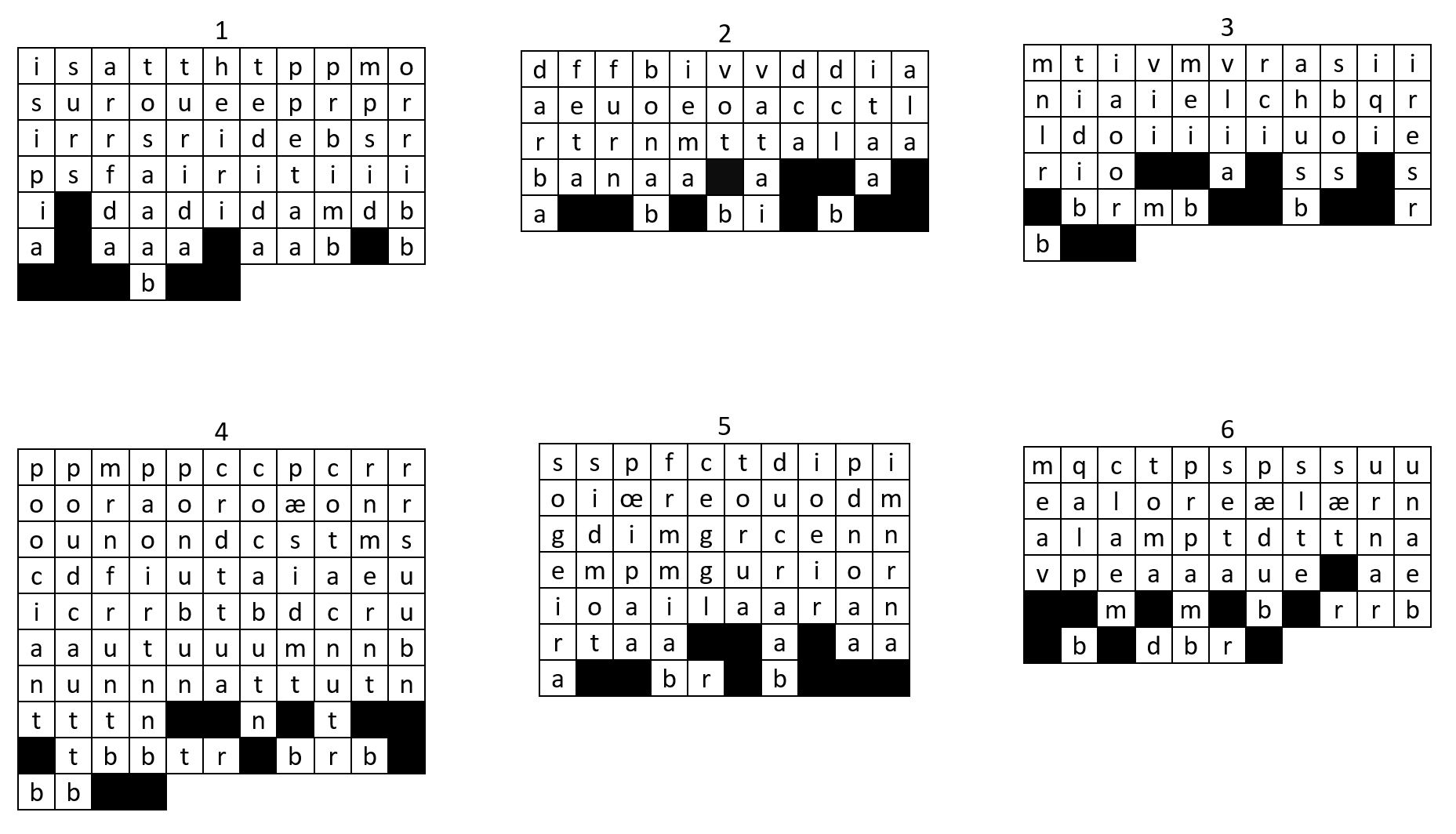}
 \caption{Tables for generating Latin verse from J. Peter, Artificial Versifying, or the School-boy’s Recreation. London, UK: John Sims, 1677.}
 \label{fig:fig5}
\end{figure}

Below are the instructions for operating the tables shown in Figure \ref{fig:fig5} to generate Latin hexameter verse [24]:
\begin{enumerate}
\item Write down any six digits from 1 to 9 to form a 6-digit number, for example: 467182. Each digit is the key to one of the tables. So, 4 is the key to Table 1, 6 is the key to Table 2, etc.
\item For the first digit (4 in the example), count up from it along the top row of Table 1 until you reach number 9. In the example, start from 5, the digit after 4, and count along the row to 9, which is the letter “t” (the fifth letter along the top row). Write that down as the first letter of your verse. [Note: Whenever you choose the digit 9, then count the first letter in the table as 1, the second 2, and so on till you reach 9, then write down that letter.]
\item Keep going along the row, counting from 1 up to 9. When you get to the end of a row, start at the beginning of the row below. For the example, from the letter “t” in the top row, count along the top row and then the row below until you reach the ninth letter (“r”). Write that down as the next letter in your verse.
\item Keep going, counting each ninth letter until you reach a black square, then stop. For the example, that should generate the remaining letters “i”, “s”, “t”, “i”, “a”, to form your first word “tristia”.
\item Do the same for the next five tables. So, for table two, the first letter in the example should be “f” (count up from 6 to 9 along the top row and write down that letter) and the remaining ones “a”, “t”, “a” forming “fata”. The complete set of words forms a line of Latin verse: “Tristia fata tibi producunt sidera prava” (meaning “Fate will produce untoward stars”).
\end{enumerate}

\end{document}